\documentclass[12pt]{article}
\usepackage{graphicx}
\usepackage{amssymb}
\usepackage{amscd}
\usepackage{amsmath}
\usepackage{appendix}

\begin{document}

{\hbox to\hsize{\hfill July 2016 }}

\bigskip \vspace{3\baselineskip}

\begin{center}
{\bf \large 
Heavy axion in asymptotically safe QCD }

\bigskip

\bigskip

{\bf Archil Kobakhidze$^{\rm a,b}$ \\ }

\smallskip

{ \small \it
$^{\rm a}$ARC Centre of Excellence for Particle Physics at the Terascale, \\
School of Physics, The University of Sydney, NSW 2006, Australia \\
$^{\rm b}$Kavli Institute for Theoretical Physics China, CAS, Beijing 100190, China\\
E-mail: archil.kobakhidze@sydney.edu.au
\\}

\bigskip
 
\bigskip

\bigskip

{\large \bf Abstract}

\end{center}
\noindent 
{\small Assuming QCD  exhibits an interacting fixed-point behaviour in the ultraviolet regime,  I argue that the axion can be substantially heavier than in the conventional case of asymptotically free QCD due to the enhanced contribution of small size instantons to its mass.  }


\vspace{1cm}


The Peecei-Quinn mechanism \cite{Peccei:1977hh} is arguably the most elegant solution to the strong CP problem. It postulates a chiral anomalous $U(1)_{\rm PQ}$ symmetry, spontaneous breaking of which results in a pseudo-Goldstone boson, known as the axion \cite{Peccei:1977hh, Weinberg:1977ma}. The axion mass $m_a$ is generated through the QCD anomaly via large size ($\gtrsim 1/\Lambda_{\rm QCD}$) instanton-mediated interactions, while the contribution of small size instantons is suppressed due to the assymptotic freedom of QCD. Also, being the pseudo-Goldstone boson, the mass of the axion ( and its couplings to the Standard Model fields) is inversely proportional to the $U(1)_{\rm PQ}$ breaking scale, $f_a$, hence $m_a\sim \Lambda_{\rm QCD}^2/f_a$. To avoid unacceptable fast colling of stars due to the axion emission together with constraints coming from various meson decays one requires $f_a\gtrsim 10^9$ GeV. Consequently, axion is a very light particle, $10^{-6}~{\rm eV} \lesssim m_a\lesssim 10^{-3}~{\rm eV}$, which interacts very feebly with ordinary matter and radiation. This is the widely accepted invisible axion scenario \cite{Kim:1979if, Zhitnitsky:1980tq}.   

It has been known for a quite sometime that high energy modification of the conventional $SU(3)_c$ colour QCD may affect the axion mass prediction \cite{Tye:1981zy}. In particular, new coloured states may drive the strong coupling constant $\alpha_s$ to increase over the range of energies, modifying the standard running towards asymptotically free regime. Alternatively, axion may couple to another confining theory with a higher confinement scale \cite{Rubakov:1997vp} or it may originate as a composite state from a high energy chiral symmetry breaking in a hypothetical sector of quarks  carrying higher than triplet colour charge \cite{Kobakhidze:2016wmv}. In all these cases physics at high energies may result in a significant increase of the axion mass.

In this brief note, I consider a scenario where $SU(3)_c$ QCD is augmented by a number of vector-like triplet quarks\footnote{The vector-like quarks are assumed to carry $U(1)_{\rm PQ}$ charge and acquire masses upon  $U(1)_{\rm PQ}$ spontaneous symmetry breaking at scale $f_a$ via Yukawa couplings to the complex scalar field responsible for the symmetry breaking. In essence, the model under consideration is just the KSVZ axion model  \cite{Kim:1979if} with multiple vector-like quarks.} such that the theory is no longer asymptotically free at high energies. Interestingly, instead of hitting the Landau pole, the theory may exhibit an interacting non-Gaussian fixed-point \cite{Holdom:2010qs}. If so, it is valid at an arbitrary high energy scale. This is known as the asymptotic safe QCD scenario.  Ignoring other Standard Model interactions, the fixed-point value of the strong coupling, $\alpha_s^{*}$, is given by \cite{Holdom:2010qs}:
\begin{equation}
\alpha_s^{*}=\frac{6\pi}{N_f+6},~~\beta(\alpha_s^{*})=0~,
\label{1}
\end{equation}
where $N_f$ is the number of vector-like quarks. This result is exact in the $N_f\to\infty$ limit and stable for finite $N_f\gtrsim 28$ \cite{Holdom:2010qs}. Using one-loop beta-functions, one can roughly estimate an upper bound on $N_f$ for a given fixed-point scale $\mu^{*}$ 
\begin{equation}
N_f\lesssim 6\pi{\alpha^{-1}_{s({\rm SM})}(\mu^*)} - 6,
\label{2}
\end{equation} 
where $\alpha_{s({\rm SM})}$ is the Standard Model strong coupling constant evaluated at $\mu^{*}$. E.g., for $\mu^{*}=M_{\rm P}\approx 2.4\cdot 10^{18}$ GeV, one finds $N_f\lesssim 929$. Thus, there is a plenty of room for the existence of a stable non-Gaussian fixed-point (\ref{1}) below Planck mass in QCD  for finite number $N_f$ of additional quarks. 

The instanton contribution to the axion mass  is typically suppressed exponentially by the familiar factor ${\rm e}^{-2\pi /\alpha_s(\mu)}$ (instanton action at a scale $\mu$) for small scale ($\Lambda_{\rm QCD}/\mu << 1$) instantons \cite{'tHooft:1976fv}. The relativily large fixed-point values of $\alpha_s^{*}$ at high energy scale $\mu^{*}$,  may avoid a strong suppression. Therefore, in what follows I will consider small instantons of size $\lesssim 1/\mu^{*}$. Also, since only vector-like quarks carry $U(1)_{\rm PQ}$ charge (besides the Peccei-Quinn scalar), they are the only ones relevant at high energies.   

Other factors involved in calculations are:  (i) ${\rm e}^{+2N_f\alpha(1/2)}$,  which comes from nonzero fermion modes and (ii) $K=\frac{2\pi^4}{\alpha_s(\mu)}{\rm e}^{-\alpha(1)-2\alpha(1/2)}$, which accounts for various zero modes \cite{'tHooft:1976fv, Bernard:1979qt},   where $\alpha(1/2)=0.145873$ and $\alpha(1)=0.443307$ have been calculated in \cite{'tHooft:1976fv}. Combining these factors and evaluating at $\mu^{*}$ we obtain: 
\begin{equation}
F(N_f) \simeq \frac{N_f^6}{1802\pi^2}{\rm e}^{-0.042 N_f}~.
\label{3}
\end{equation}

Next, for the sake of simplicity, let me assume that all the $N_f$ vector-like quarks have approximately the same mass $m_f\lesssim \mu^{*}$.  If the fermion zero modes are closed up by mass term insertions in the 'tHooft vertex, the contribution of small scale instantons to the axion mass squared can be readily computed as: 
 \begin{equation}
m^2_a=\frac{F(N_f)}{N_f-4}\left(\frac{m_f}{\mu^*}\right)^{N_f}\frac{\mu^{*4}}{f_a^2}~, 
\label{4}
\end{equation}
where I have used again 1-loop approximation to evaluate the factor $(m_f/\mu^*)^{N_f}$:
\begin{equation}
\left(\frac{m_f}{\mu^*}\right)^{N_f}\approx {\rm epx}\left[-\frac{3\pi}{\alpha_{s({\rm SM})}(\mu^*)}+\frac{N_f}{2}+3\right]~.
\label{5}
\end{equation}
Recall, $N_f$ is subject to the constraint (\ref{2}).    

Some numerics are presented in Table 1, from which one can see that contribution from small instantons to the axion mass can dominate over the standard large instanton contribution by many orders of magnitude. Depending on parameters, the axion mass can range from astrophysically safe hundreds of keVs to hundreds of TeVs or higher. If one feels uncomfortable with large multiplicity of vector-like quarks, an alternative scenario with fewer quarks in high-colour representation of $SU(3)_c$ \cite{Kobakhidze:2016wmv} may also be considered. Let me also note that a large multiplicity of vector-like quarks naturally emerges in theories with extra dimensions as the Kaluza-Klein excitations of the Standard Model quarks.   
\begin{table}[t]
\label{table1}
\begin{center}
\begin{tabular}{|c|c|c|c|c|}
\hline 
$N_f$ & 30 & 150 & 200 & 250 \\ 
\hline 
$\mu^{*}=10^5$ GeV & $5.0\cdot 10^{-29}$ & $2.6\cdot 10^{-15}$ & $9.2\cdot 10^{-9}$ & $1.1\cdot 10^{-4}$ \\ 
\hline 
\hline
$N_f$ & 150 & 300 & 400 & 500 \\ 
\hline 
$\mu^{*}=10^{10}$ GeV & $2.1\cdot 10^{-41}$ & $9.0\cdot 10^{-26}$ & $1.8\cdot 10^{-15}$ & 0.003  \\ 
\hline
\end{tabular}
\end{center}
\vspace{-0.3cm}
{\small \caption{Contributions of small instantons to the axion mass squared measured in units of $\mu^{*2}/f_a$ for various $N_f$ and $\mu^{*}$.} }
\end{table}

Eq. (\ref{4}) may not be the end of the story. It is conceivable to think that the vector-like quarks in the strongly coupled regime at high energies may form condensates\footnote{If so, a more economical scenario without  elementary scalars may work, with axion being a composite state \cite{Kobakhidze:2016wmv}.}. The fermion zero modes then can be tied up through these condensates. The contribution to the axion mass squared now reads:        
\begin{equation}
m_a^2=F(N_f)\int_0^{1/\mu^*}\frac{d\rho}{\rho^5}\left(\Lambda \rho\right)^3=
\frac{F(N_f)}{3N_f-4}\left(\frac{\Lambda}{\mu^*}\right)^{3N_f}\frac{\mu^{*4}}{f_a^2}~,
\label{6}
\end{equation}
 where $\Lambda$ is the scale of vector-like quark condensation. It is reasonable to expect that  $\Lambda \sim \mu^{*}$ and we have no strong suppression unlike the previous estimation in Eq. (\ref{4}). 
 
The lesson from this exercise is that in modified QCD theories the prediction for axion mass may crucially depend of the ultraviolet behaviour of the theory. I have considered explicitly the scenario of assymptotic safe QCD and found that a heavy axion can be accommodated thanks to the large contribution of small size instantons to the axion mass. Obviously, phenomenology and cosmology of such heavy axions are entirely different from the case of the standard invisible axion. It would be interesting to perform a detailed study in this direction. 

The solution to the strong CP problem may be compromised in some axion models with additional CP violation and potentially large contribution from small scale instantons. Therefore, this contribution should be be carefully evaluated in such cases.  Finally,  let me also note that the large amplitude axion potential generated by small instantons may cure some problems of the cosmological relaxation mechanism for the Higgs mass \cite{Graham:2015cka}.     

\paragraph{Acknowledgement.} I gratefully acknowledge partial support by the Australian Research Council and Shota Rustaveli National Science Foundation (grant DI/12/6-200/13). This work was finalised during my visit at KITPC.


\small
\begin{thebibliography}{999}
\bibitem{Peccei:1977hh} 
  R.~D.~Peccei and H.~R.~Quinn,
  Phys.\ Rev.\ Lett.\  {\bf 38}, 1440 (1977);
  Phys.\ Rev.\ D {\bf 16}, 1791 (1977).

\bibitem{Weinberg:1977ma} 
  S.~Weinberg,
  Phys.\ Rev.\ Lett.\  {\bf 40}, 223 (1978); 
  F.~Wilczek,
  Phys.\ Rev.\ Lett.\  {\bf 40}, 279 (1978).
  
\bibitem{Kim:1979if} 
  J.~E.~Kim,
  Phys.\ Rev.\ Lett.\  {\bf 43}, 103 (1979); 
  M.~A.~Shifman, A.~I.~Vainshtein and V.~I.~Zakharov,
  Nucl.\ Phys.\ B {\bf 166}, 493 (1980).

\bibitem{Zhitnitsky:1980tq} 
  A.~R.~Zhitnitsky,
  Sov.\ J.\ Nucl.\ Phys.\  {\bf 31}, 260 (1980)
  [Yad.\ Fiz.\  {\bf 31}, 497 (1980)]; 
  M.~Dine, W.~Fischler and M.~Srednicki,
  Phys.\ Lett.\ B {\bf 104}, 199 (1981).
  
\bibitem{Tye:1981zy} 
  S.~H.~H.~Tye,
  Phys.\ Rev.\ Lett.\  {\bf 47}, 1035 (1981);
  B.~Holdom and M.~E.~Peskin,
  Nucl.\ Phys.\ B {\bf 208}, 397 (1982);
  M.~Dine and N.~Seiberg,
  Nucl.\ Phys.\ B {\bf 273}, 109 (1986);
  J.~M.~Flynn and L.~Randall,
  Nucl.\ Phys.\ B {\bf 293}, 731 (1987).
  
\bibitem{Rubakov:1997vp} 
  V.~A.~Rubakov,
  JETP Lett.\  {\bf 65}, 621 (1997)
  [hep-ph/9703409]; 
  Z.~Berezhiani, L.~Gianfagna and M.~Giannotti,
  Phys.\ Lett.\ B {\bf 500}, 286 (2001)
  [hep-ph/0009290]. 
Some more recent works are: 
  A.~Hook,
  Phys.\ Rev.\ Lett.\  {\bf 114}, no. 14, 141801 (2015)
  [arXiv:1411.3325 [hep-ph]];
  H.~Fukuda, K.~Harigaya, M.~Ibe and T.~T.~Yanagida,
  Phys.\ Rev.\ D {\bf 92}, no. 1, 015021 (2015)
  [arXiv:1504.06084 [hep-ph]].
  T.~Gherghetta, N.~Nagata and M.~Shifman,
  Phys.\ Rev.\ D {\bf 93}, no. 11, 115010 (2016)
  [arXiv:1604.01127 [hep-ph]].
  
\bibitem{Kobakhidze:2016wmv} 
  A.~Kobakhidze,
  arXiv:1602.06363 [hep-ph]; 
  N.~D.~Barrie, A.~Kobakhidze, M.~Talia and L.~Wu,
  Phys.\ Lett.\ B {\bf 755}, 343 (2016)
  [arXiv:1602.00475 [hep-ph]].

\bibitem{Holdom:2010qs} 
  B.~Holdom,
  Phys.\ Lett.\ B {\bf 694}, 74 (2011)
  [arXiv:1006.2119 [hep-ph]]; 
  D.~F.~Litim and F.~Sannino,
  JHEP {\bf 1412}, 178 (2014)
  [arXiv:1406.2337 [hep-th]].
  
\bibitem{'tHooft:1976fv} 
  G.~'t Hooft,
  Phys.\ Rev.\ D {\bf 14}, 3432 (1976)
  Erratum: [Phys.\ Rev.\ D {\bf 18}, 2199 (1978)]; 
  Phys.\ Rept.\  {\bf 142}, 357 (1986).
  
\bibitem{Bernard:1979qt} 
  C.~W.~Bernard,
  Phys.\ Rev.\ D {\bf 19}, 3013 (1979).
  
\bibitem{Graham:2015cka} 
  P.~W.~Graham, D.~E.~Kaplan and S.~Rajendran,
  Phys.\ Rev.\ Lett.\  {\bf 115}, no. 22, 221801 (2015)
  [arXiv:1504.07551 [hep-ph]];
  R.~S.~Gupta, Z.~Komargodski, G.~Perez and L.~Ubaldi,
  JHEP {\bf 1602}, 166 (2016)
  [arXiv:1509.00047 [hep-ph]].

\end{thebibliography}
\end{document}